\documentclass[dvips,12pt]{article}
\usepackage{graphicx}
\usepackage{repko}
\def\ds{\displaystyle}

\def\sb{\mbox{\rule{0pt}{11pt}}}

\def\al{\alpha}
\def\be{\beta}
\def\ga{\gamma}
\def\de{\delta}
\def\la{\lambda}
\def\ro{\rho}
\def\ep{\varepsilon}

\newfont{\ex}{cmr10}
\begin{document}
\title{\hfill {\small UTEXAS-HEP-00-2} \\ [-8pt]
\hfill  {\small MSUHEP-00310} \\ [-8pt]
\hfill \\
{\large\bf Photons, neutrinos and large compact space dimensions}}
\author{\normalsize Duane A. Dicus$^1$, Kristin Kovner$^2$ and Wayne W. Repko$^3$ \\
{\small\it $^1$Center for Particle Physics and Department of Physics,
University of Texas, Austin, Texas 78712} \\
{\small\it $^2$Schreiber High School, Port Washington, NY 11050}\\
{\small\it $^3$Department of
Physics and Astronomy, Michigan State University, East Lansing, Michigan
48824}}
\date{\normalsize\today}
\maketitle
\begin{abstract}
We compute the contribution of Kaluza-Klein graviton exchange to the cross
section for photon-neutrino scattering. Unlike the usual situation where the
virtual graviton exchange represents a small correction to a leading order
electroweak or strong amplitude, in this case the graviton contribution is
of the same order as the electroweak amplitude, or somewhat larger. Inclusion of the
graviton contribution is not sufficient to allow high energy neutrinos to scatter
from relic neutrinos in processes such as $\nu\bar{\nu}\to\gamma\gamma$, but the
photon-neutrino decoupling temperature is substantially reduced.
\end{abstract}

\section{Introduction}

The $2\to 2$ processes $\ga\nu\to\ga\nu$, $\ga\ga\to\nu\bar{\nu}$ and
$\nu\bar{\nu}\to\ga\ga$ are of potential interest in astrophysics. However,
because of the vector-axial-vector nature of the weak coupling, the leading
term in these cross sections for these processes with massless neutrinos,
nominally of order $G_F^2\al^2\omega^2$, vanishes due to Yang's theorem
\cite{GellMann,Yang}. In the limit that the photon energy $\omega<m_e$, where
$m_e$ is the electron mass, these cross sections can be shown to be of order
$G_F^2\al^2\omega^2\left(\omega/m_W\right)^4$ \cite{Levine,LS,DR93}, and, in
the annihilation channels at least, this $\omega^6$ behavior persists to center
of mass energies $\sqrt{s}\sim 2m_W$, where $m_W$ is the mass of the $W$-boson
\cite{ADDR}.

The $\omega^6$ behavior of the photon-neutrino cross sections can be understood
in terms of an effective Lagrangian  of the form
\begin{equation}\label{leff}
{\mathcal L}^{\mathrm{SM}}_{\mathrm{eff}} =
\frac{1}{32\pi}\frac{g^2\al}{m_W^4}A\left[\bar{\psi}\ga_{\nu}(1+\ga_5)
(\partial_{\mu}\psi)-(\partial_{\mu}\bar{\psi})\ga_{\nu}(1+\ga_5)\psi\right]
F_{\mu\la}F_{\nu\la}\,,
\end{equation}
where $g$ is the electroweak gauge coupling, $\psi$ is the neutrino field and
$F_{\mu\nu}$ is the electromagnetic field tensor. In Eq.\,(\ref{leff}), $A$ is
\begin{equation}
A = \left[\frac{4}{3}\ln\left(\frac{m_W^2}{m_e^2}\right)+1\right]
\end{equation}
in the low energy limit $\omega<m_e$, and $A$ is obtained by fitting the
numerical calculation of the cross section for $\omega>m_e$ \cite{ADDR}. Since
${\mathcal L}^{\mathrm{SM}}_{\mathrm{eff}}$ is a dimension 8 operator, it
follows that the center of mass cross sections will behave as $\omega^6$.
Another property of Eq.\,(\ref{leff}) is that the scattered photons in the
channel $\ga\nu\to\ga\nu$ are circularly polarized {\em in leading order}
\cite{DR93} due to the parity violating terms in ${\mathcal L}_{\mathrm{eff}}$.
There is no linear polarization in this channel.

Because the scale of ${\mathcal L}^{\mathrm{SM}}_{\mathrm{eff}}$ is $m_W$, a
typical photon-neutrino cross section is quite small, so small that a high
energy neutrino beam is not attenuated by interactions with the present density
of relic neutrinos via the process $\nu\bar{\nu}\to\ga\ga$ \cite{ADDR}. In the
early universe, the photons and neutrinos decouple at a temperature
$T\sim$\,1.6 GeV, or about one microsecond after the Big Bang. If this
temperature were a factor of 10 lower, {\em i.e.}
$T\leq\Lambda_{\mathrm{QCD}}$, one might be justified in speculating that some
remnant of the circular polarization mentioned above could be retained in the
cosmic microwave background radiation and at this would provide evidence for
the relic neutrino background.

Lowering the decoupling temperature necessitates increasing the cross section,
$\sigma(\nu\bar{\nu}\to\ga\ga)$, or changing the dependence of the age of the universe,
$t$, on the temperature $T$. The latter seems unlikely, since the $t\sim T^{-2}$
radiation dominated behavior of the early universe is insensitive changes such as
including a non-vanishing cosmological constant. On the other hand, new interactions
could increase $\sigma(\nu\bar{\nu}\to\ga\ga)$, provided they involve the exchange of
particles with spin $\neq$\,1. A new interaction of this type is provided by the recent
proposal that the compact dimensions of string theory are sufficiently large to make the
effective gravitational scale $\Lambda$ of order a TeV rather than the usual
$M_P=1.2\times 10^{19}$\,GeV Planck scale \cite{model,witten,GRW,peskin,HLZ}. That this
new gravitational interaction will make a significant correction to the standard model
photon-neutrino cross sections can be seen by rewriting Eq.\,(\ref{leff}) in the form
\begin{equation}\label{lsm}
{\mathcal L}^{\mathrm{SM}}_{\mathrm{eff}} =
\frac{1}{8\pi}\frac{g^2\al}{m_W^4}AT^{\nu}_{\al\be}T^{\ga}_{\al\be}\,,
\end{equation}
where $T^{\nu}_{\al\be}$ and $T^{\ga}_{\al\be}$ are the symmetrical
energy-momentum tensors of the neutrinos and the photons. Explicitly, we have
\begin{eqnarray}
T^{\nu}_{\al\be} & = & \frac{1}{8}
\left[\bar{\psi}\ga_{\al}(1+\ga_5)(\partial_{\be}\psi) +
\bar{\psi}\ga_{\be}(1+\ga_5)(\partial_{\al}\psi) \right. \nonumber           \\
&  &\left.-(\partial_{\be}\bar{\psi})\ga_{\al}(1+\ga_5)\psi -
(\partial_{\al}\bar{\psi})\ga_{\be}(1+\ga_5)\psi\right]\,,  \\[4pt]
T^{\ga}_{\al\be} & = &
F_{\al\la}F_{\be\la}-\frac{1}{4}\de_{\al\be}F_{\la\ro}F_{\la\ro}\,.
\end{eqnarray}

In the next section, we show that an effective interaction which is the product of
energy-momentum tensors also arises when the spin 2 graviton is exchanged between photons
and neutrinos. This is followed by the calculation of
$\sigma(\nu\bar{\nu}\to\gamma\gamma)$ and a discussion of the resulting astrophysical
implications.

\section{Graviton exchange between photons and neutrinos}

According to models in which only the graviton (${\mathcal G}$) propagates in
the additional $n$ compact dimensions of a $D=4+n$ dimensional manifold, the
compact spatial dimension $R$ is related to $\Lambda$ and $M_P$ as \cite{model}
\begin{equation}\label{scale}
\Lambda^{n+2}R^n\sim M_p^2/4\pi\,.
\end{equation}
The graviton's propagation in all $D$ dimensions implies the existence of a
tower of spin 2 particles in ordinary space-time, whose masses are given by
$m^2_{\vec{n}}=\vec{n}^2/R^2$, where $\vec{n} = (n_1,n_2,\ldots,n_n)$ and the
$n_i$ are integers. Furthermore, in these models the interaction between the
spin 2 graviton, ${\mathcal G}_{\mu\nu}$, and any standard model field has the
universal form
\begin{equation}\label{lgeff}
{\mathcal L}^{\mathcal G}_{\mathrm{eff}} = -\frac{\kappa}{2}T_{\mu\nu}{\mathcal
G}_{\mu\nu}\,,
\end{equation}
where $T_{\mu\nu}$ is the energy-momentum tensor of the standard model field.

Given the effective coupling, Eq.\,(\ref{lgeff}), it is a simple matter to calculate the
$2\to 2$ amplitudes \cite{KC}. In the annihilation channels ($\ga\ga\to\nu\bar{\nu}$ and
$\nu\bar{\nu}\to\ga\ga)$, the amplitude for a particular $m_{\vec{n}}$ has the form
\begin{equation}\label{amp1}
{\mathcal A}_{\vec{n}} =
\frac{\kappa^2}{4}T^{\nu}_{\al\be}(p_1,p_2)\,\frac{{\mathcal
P}_{\al\be\la\ro}(k_1+k_2)}{m^2_{\vec{n}}-s-i\ep}\,T^{\ga}_{\la\ro}(k_1,k_2)\,,
\end{equation}
with $s=-(k_1+k_2)^2$. Here ${\mathcal P}_{\al\be\la\ro}$ is the spin 2
projection operator
\begin{equation}
{\mathcal P}_{\al\be\la\ro}(k)=\frac{1}{2}\left(d_{\al\la}(k)d_{\be\ro}(k) +
d_{\al\ro}(k)d_{\be\la}(k)-\frac{2}{3}d_{\al\be}(k)d_{\la\ro}(k)\right)\,,
\end{equation}
with
\begin{equation}
d_{\al\be}(k) = \de_{\al\be}+\frac{1}{m^2_{\vec{n}}}k_{\al}k_{\be}\,.
\end{equation}
Since the energy-momentum tensors are conserved, symmetrical and, in this case,
traceless, Eq.\,(\ref{amp1}) reduces to
\begin{equation}\label{amp2}
{\mathcal A}_{\vec{n}} =
\frac{\kappa^2}{4}T^{\nu}_{\al\be}(p_1,p_2)\,\frac{1}{m^2_{\vec{n}}-s-i\ep}
\,T^{\ga}_{\al\be}(k_1,k_2)\,.
\end{equation}
To complete the calculation of the amplitude, it is necessary to sum ${\mathcal
A}_{\vec{n}}$ over the values of $m^2_{\vec{n}}$. This is done by replacing the
sum with an integral over the number density $d\mathcal{N}$ given by
\cite{GRW,HLZ}
\begin{equation}
d{\mathcal{N}} = \frac{1}{2}\Omega_nR^n\,(m^2)^{(n-2)/2}dm^2\,,
\end{equation}
where $\Omega_n$ is the surface area of an $n$-dimensional sphere. If the
integral over $dm^2$ is cut off at $\Lambda^2$, we find as leading terms
\cite{HLZ}
\begin{equation}
\int_0^{\Lambda^2}\frac{d{\mathcal{N}}}{m^2-s-i\ep}=
        \frac{1}{2}\Omega_nR^n\Lambda^{n-2}
        \left\{
        \begin{array}{ccc}
        \ln\left(\frac{\ds\Lambda^2}{\ds s}\right)& \mathrm{if} &
        n=2  \\ [8pt]
        \frac{\ds 2}{\ds\sb (n-2)} & \mathrm{if} & n>2
       \end{array}
       \right. = \frac{1}{2}\Omega_nR^n\Lambda^{n-2}I_n(\Lambda,s)\,.
\end{equation}
Apart from a $\ln(\Lambda^2/s)$ term when $n=2$, all values of $n$ have the
same $\Lambda^{n-2}$ dependence.
If we then take the specific realization of Eq.\,(\ref{scale}) for the scale
parameter $\Lambda$ to be
\begin{equation}
\Omega_n\Lambda^{n+2}R^n = M_P^2\,,
\end{equation}
the summed version of Eq.\,(\ref{amp2}) is
\begin{equation}\label{lgrav}
\mathcal{A}_{\mathcal{G}} =
\frac{4\pi}{\Lambda^4}I_n(\Lambda,s)T^{\nu}_{\al\be}(p_1,p_2)\,T^{\ga}_{\al\be}(k_1,k_2)\,,
\end{equation}
where we have used $\kappa^2 = 32\pi/M_P^2$ \cite{GRW}.

\section{\mbox{\boldmath $\nu\bar{\nu}\protect\to\ga\ga$} cross section including
graviton exchange}

\begin{figure}[h]
\hfill\includegraphics[height=2.4in]{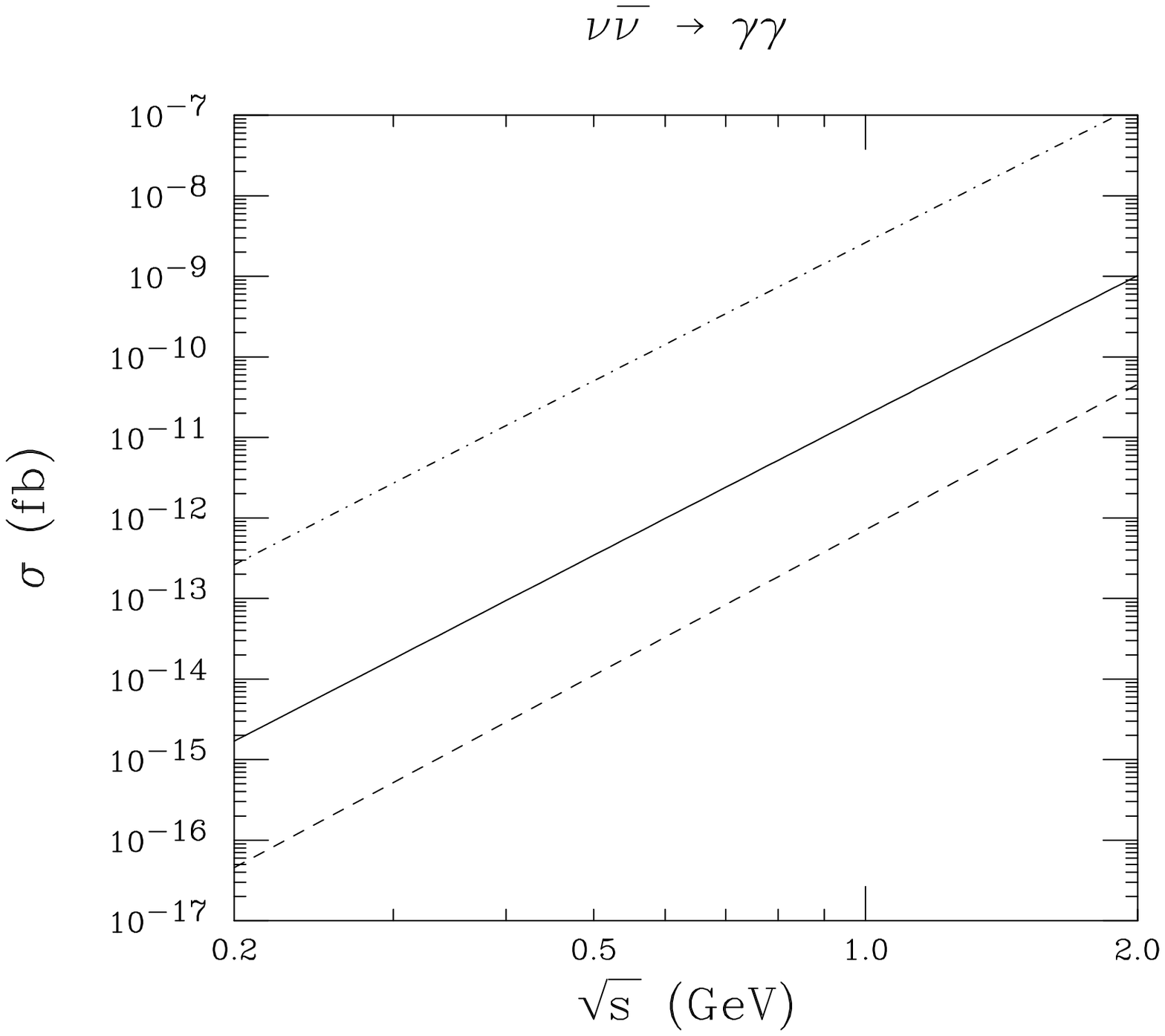}%
\hfill\includegraphics[height=2.4in]{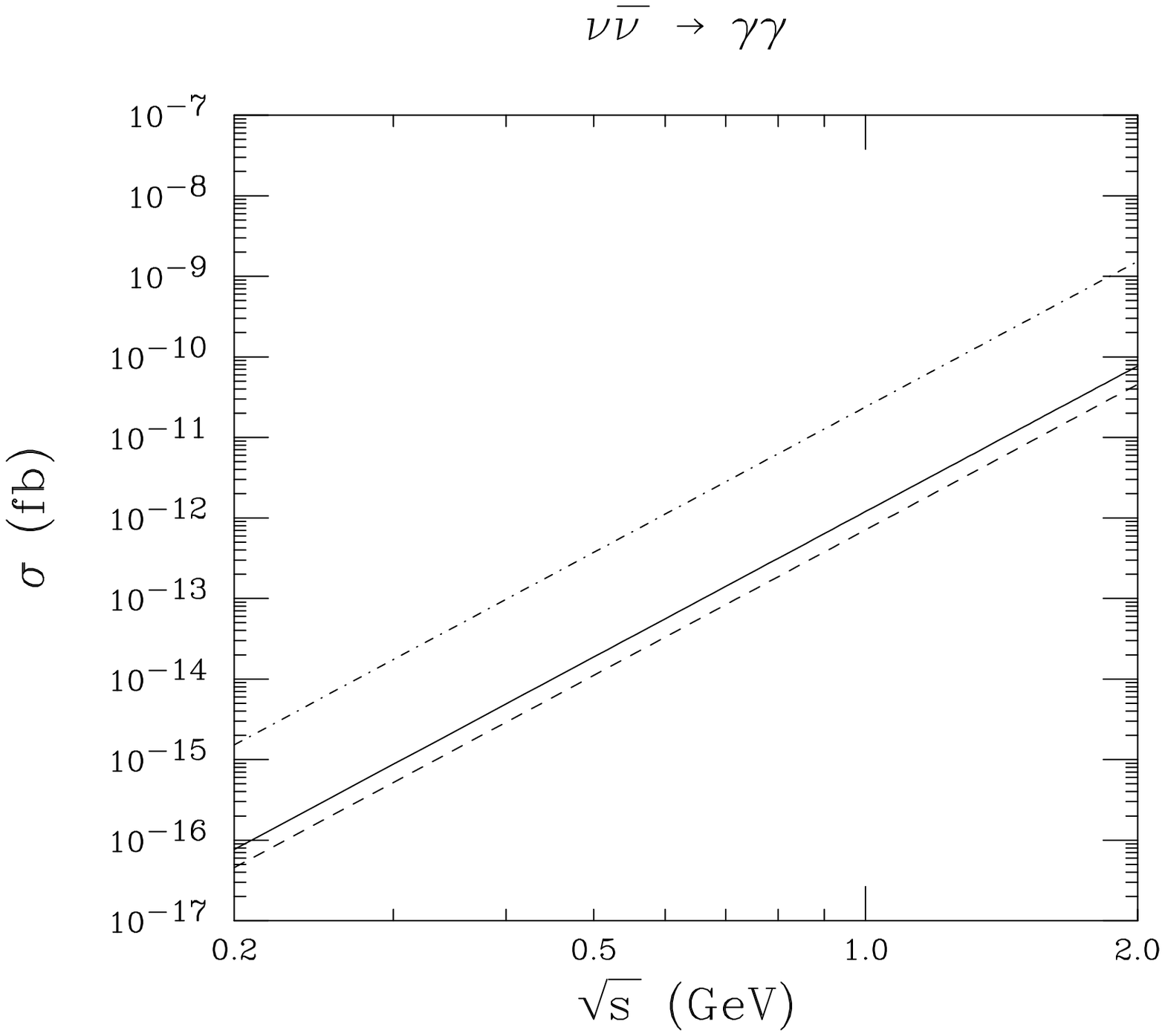}\hspace*{0pt\hfill}
\caption{\footnotesize The total cross section
$\sigma(\nu\bar{\nu}\protect\to\ga\ga)$ is shown. The left panel is the $n=2$
result and the right panel is the $n=4$ result. In each panel, the solid line
corresponds to $\Lambda=1\,\mathrm{TeV}$, the dot-dash line to $\Lambda =
.5\,\mathrm{TeV}$ and the dashed line to $\Lambda=10\,\mathrm{TeV}$. The
$\Lambda=10\,\mathrm{TeV}$ curve is identical to the standard model result.
\label{fig:sig}}
\end{figure}

 Using Eq.\,(\ref{lsm}), the Standard Model amplitude for the
annihilation processes is
\begin{equation}
\mathcal{A}^{\mathrm{SM}}=\frac{1}{8\pi}\frac{g^2\al}{m_W^4}A
T^{\nu}_{\al\be}(p_1,p_2)\,T^{\ga}_{\al\be}(k_1,k_2)\,,
\end{equation}
which, in view of Eq.\,(\ref{lgrav}), leads to the total amplitude
\begin{equation}\label{ampt}
\mathcal{A} = \left(\frac{1}{8\pi}\frac{g^2\al}{m_W^4}A +
\frac{4\pi}{\Lambda^4}I_n(\Lambda,s)\right)T^{\nu}_{\al\be}(p_1,p_2)\,T^{\ga}_{\al\be}(k_1,k_2)
\,.
\end{equation}
If the photon helicities are denoted by $\la_1$ and $\la_2$, the product of
energy-momentum tensors in Eq.\,(\ref{ampt}) is given by
\begin{eqnarray}
T^{\nu}_{\al\be}(p_1,p_2)\,T^{\ga}_{\al\be}(k_1,k_2) & = &\frac{1}{4}
\sin\theta\left[st(1-\la_1\la_2)+\frac{1}{2}s^2(1-\la_1)(1+\la_2)\right]\,, \\
& = &\frac{1}{4}\mathcal{M}_{\la_1\la_2}\,,
\end{eqnarray}
where $t=-(p_1-k_1)^2$, and $\theta$ is the scattering angle in the center of
mass. The differential cross section for $\nu\bar{\nu}\to\ga\ga$ can then be
calculated using
\begin{equation}
\frac{d\sigma}{dz}=\frac{1}{32\pi s}
\sum_{\la_1\la_2}|\mathcal{A}_{\la_1\la_2}|^2\,,
\end{equation}
with $z=\cos\theta$, and
\begin{equation}
\mathcal{A}_{\la_1\la_2} = \left(\frac{1}{32\pi}\frac{g^2\al}{m_W^4}A +
\frac{\pi}{\Lambda^4}I_n(\Lambda,s)\right)\mathcal{M}_{\la_1\la_2}\,.
\end{equation}
Summing over the helicities gives
\begin{equation}
\frac{d\sigma}{dz}=\frac{1}{16\pi}\left(\frac{1}{32\pi}\frac{g^2\al}{m_W^4}A +
\frac{\pi}{\Lambda^4}I_n(\Lambda,s)\right)^2s^3(1-z^4)\,,
\end{equation}
which leads to the total cross section
\begin{eqnarray}
\sigma(\nu\bar{\nu}\to\ga\ga)& = &\frac{1}{2!}\int_{-1}^{1}dz\frac{d\sigma}{dz}
\\ [4pt]
& = &\frac{s^3}{20\pi}\left(\frac{1}{32\pi}\frac{g^2\al}{m_W^4}A +
\frac{\pi}{\Lambda^4}I_n(\Lambda,s)\right)^2\,.
\end{eqnarray}
Using the value $A=14.4$ \cite{ADDR} and expressing $\Lambda$ in TeV, we find
\begin{equation}
\sigma(\nu\bar{\nu}\to\ga\ga) = \frac{1}{8}\frac{s^3}{m_e^6}\left(1 +
.3\frac{I_n(\Lambda,s)} {\Lambda^4}\right)^2\times 10^{-31}\;\mathrm{fb} \,.
\end{equation}
The cross section is shown in Fig.\,\ref{fig:sig} for the cases $n=2$ and $n=4$
with $\Lambda = .5,\, 1\,\mathrm{and}\,10$\,TeV. Although the logarithmic
variation in the $n=2$ case is scarcely detectable in the range
$.2\,\mathrm{GeV}\leq\sqrt{s}\leq 2\,\mathrm{GeV}$, its presence in the
coefficient of $\Lambda^{-4}$ makes the effect of the extra dimensions largest
for this case.

\section{Discussion and conclusions}

The possibility of a high energy neutrino scattering from the current relic
neutrino background is not materially  enhanced by the inclusion of the effects
of gravitons propagating in compact dimensions. Neglecting the electroweak
contribution to the $\nu\bar{\nu}\to\gamma\gamma$ cross section, which is known
to be too small \cite{ADDR} to produce any scattering, the condition
$\sigma_{\nu\bar{\nu}\to\gamma\gamma}n_{\nu}ct_0=1$ for at least one scattering,
gives
\begin{equation}
\frac{\pi}{20}\frac{s^3}{\Lambda^8}\ln^2\left(\frac{\Lambda^2}{s}\right)n_{\nu}ct_0=1\,
\end{equation}
or, using the relic neutrino density $n_{\nu}=56\mathrm{cm}^{-3}$, and the age
of the universe $t_0=15\times 10^9$\,years, the condition can be written
\begin{equation}\label{scat}
x^6\ln^2(x^2)=0.0207\tilde{\Lambda}^2\,,
\end{equation}
with $x=\sqrt{s}/\Lambda$ and $\tilde{\Lambda}$ in GeV. For a 1 TeV scale, the
solution to Eq.\,(\ref{scat}) is $x=3.78$, giving $\sqrt{s}=3.78$\,TeV, which
is beyond the range of validity of the effective theory.

The additional contribution to the cross section from gravition exchange will affect the
decoupling temperature. The temperature at which the reaction
$\nu\bar{\nu}\to\gamma\gamma$ ceases to occur can be determined from the reaction rate
per unit volume
\begin{equation} \label{rho}
\rho = \frac{1}{(2\pi)^6}\int\frac{d^3p_1}{e^{E_1/T} + 1} \int\frac{d^3p_2}{e^{E_2/T} +
1}\sigma|\vec{v}|\,,
\end{equation}
where $\vec{p}_1$ and $\vec{p}_2$ are the neutrino and antineutrino momenta, $E_1$ and
$E_2$ their energies, $|\vec{v}|$ is the flux and $T$ the temperature. Using the
invariance of $\sigma E_1E_2|\vec{v}|$, the relationship between $\sigma |\vec{v}|$ in
the center of mass frame and any other frame is
\begin{equation}\label{sigv}
\sigma |\vec{v}| = \sigma_{CM}\frac{2E_{\rm CM}^2}{E_1E_2}\,,
\end{equation}
which gives
\begin{equation}
\sigma |\vec{v}| = \frac{s^4}{16E_1E_2m_e^6}\left(1+\frac{.3}{\Lambda^4}
\ln\left(\frac{\Lambda^2}{s}\right)\right)^2\times 10^{-70}\mathrm{cm}^2\,,
\end{equation}
for $n=2$ and $\Lambda$ in TeV. Taking $s=4E_1E_2\sin^2(\theta_{12}/2)$\,, where $\theta_{12}$
is the angle between the incoming neutrinos, the angular integrations in Eq.\,(\ref{rho})
result in the integrand
\begin{equation}\label{angint}
\int d\Omega_1 d\Omega_{12} = \frac{(4\pi)^2}{5}\left[\left(1+\frac{.3}{\Lambda^4}
\ln\left(\frac{\Lambda^2}{4E_1E_2}\right)\right)^2
+\frac{.12}{\Lambda^4}\left(1+\frac{.3}{\Lambda^4}\ln\left(\frac{\Lambda^2}
{4E_1E_2}\right)\right)+.08\left(\frac{.3}{\Lambda^4}\right)^2\right]\,.
\end{equation}
To estimate the decoupling temperature, the last two terms on the right in Eq.\,(\ref{angint}),
which are suppressed relative to the first by small numerical factors, can be neglected. The
reaction rate per unit volume is then given by
\begin{equation}\label{rho1}
\rho  =  \frac{6.4\times 10^{-69}\mathrm{cm}^2}{5(2\pi)^4}\frac{T^{12}}{m_e^6}\int_0^{\infty}dx
\frac{x^5}{e^x+1}\int_0^{\infty}dy\frac{y^5}{e^y+1}
\left[1+\frac{.3}{\Lambda^4}\left(\ln\left(\frac{\Lambda^2}
{4T^2}\right)-\ln x-\ln y\right)\right]^2\,.
\end{equation}
The $\ln x$ and $\ln y$ terms in Eq.\,(\ref{rho1}) result in contributions which are
small relative to the remaining terms. Omitting these terms gives
\begin{equation}\label{rho2}
\rho = \frac{6.4\times 10^{-69}\mathrm{cm}^2}{5(2\pi)^4}\frac{T^{12}}{m_e^6}
\left[1+\frac{.3}{\Lambda^4}\ln\left(\frac{\Lambda^2}{4T^2}\right)\right]^2
\left[\frac{31}{32}\Gamma(6)\zeta(6)\right]^2\,,
\end{equation}
where $\zeta(z)$ is the Riemann Zeta function. The interaction rate $R$ is obtained by
dividing Eq.\,(\ref{rho2}) by the neutrino density $n_{\nu}=3\zeta(3)T^3/4\pi^2$, giving
\begin{equation}
R=7.3\times 10^{-24}T_{10}^9\left[1+\frac{.3}{\Lambda^4}\ln\left(\frac{10^{12}\Lambda^2}
{3T_{10}^2}\right)\right]^2\mathrm{sec}^{-1}\,,
\end{equation}
with $T_{10}=T/10^{10}K$. Multiplying $R$ by the age of the universe, $t=2T_{10}^{-2}$\, sec,
the condition for a single interaction to occur is
\begin{equation}
\frac{T_{10}}{1828}\left[1+\frac{.3}{\Lambda^4}\ln\left(\frac{10^{12}\Lambda^2}{3T_{10}^2}
\right)\right]^{2/7}= 1\,.
\end{equation}
\begin{figure}
\centering\includegraphics[height=2.4in]{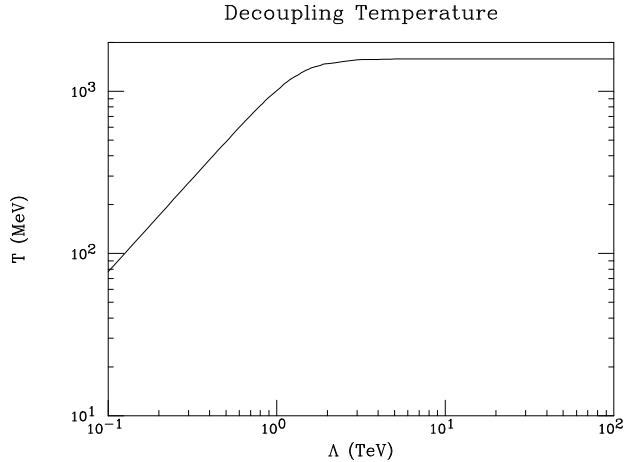}
\caption{\footnotesize The decoupling temperature is shown as a function of the effective
gravitational scale $\Lambda$. \label{decoup}}
\end{figure}
The solution to this equation is shown in Fig.\,\ref{decoup}. While there is a
substantial correction to the Standard Model decoupling temperature, a decoupling
temperature of a few hundred MeV is only possible for $\Lambda\sim 250-300\,$GeV, which
is unrealistically low. The decoupling temperature for a 1 TeV scale is 1 GeV, down from
the Standard Model result of 1.6 GeV. Thus, a mechanism for lowering the photon-neutrino
decouping temperature below $\Lambda_{\mathrm{QCD}}$ remains elusive.

\begin{center}
\section*{Acknowledgement}
\end{center}
One of us (K.K.) wishes to thank her fellow participants in the Michigan State University
High School Honors Science Program for numerous helpful conversations. This research was
supported in part by the National Science Foundation under grant PHY-9802439 and by the
Department of Energy under Contract No. DE-FG13-93ER40757.
\newpage
\begin{center}
\section*{References}
\end{center}

\end{document}